\documentclass[12pt,preprint]{aastex}
\begin{document}
\title{{\it GALEX}, Optical and IR Light Curves of MQ Dra: UV Excesses at Low Accretion Rates}

\author{Paula Szkody\altaffilmark{1}, 
Albert P. Linnell\altaffilmark{1},
Ryan K. Campbell\altaffilmark{2},
Richard M. Plotkin\altaffilmark{1},
Thomas E. Harrison\altaffilmark{2},
Jon Holtzman\altaffilmark{2},
Mark Seibert\altaffilmark{3},
Steve B. Howell\altaffilmark{4}}
\altaffiltext{1}{Department of Astronomy, University of Washington,
Box 351580,
Seattle, WA 98195, szkody@astro.washington.edu,plotkin@astro.washington.edu,linnell@astro.washington.edu}
\altaffiltext{2}{Department of Astronomy, New Mexico State University, Box 30001, MSC 4500, Las Cruces, NM 88003, cryan@nmsu.edu,tharriso@nmsu.edu,holtz@nmsu.edu}
\altaffiltext{3}{Observatories Carnegie Institute of Washington, 813 Santa Barbara St., Pasadena CA 91101, mseibert@ociw.edu}
\altaffiltext{4}{National Optical Astronomy Observatories,
950 N. Cherry Avenue, Tucson, AZ 85726, howell@noao.edu}

\begin{abstract}
Ultraviolet light curves constructed from NUV and FUV detectors on {\it GALEX}
reveal large amplitude variations during the orbital period of the Low Accretion Rate
Polar MQ Dra (SDSSJ1553+55). This unexpected variation from a UV source is similar
to that seen and discussed 
in the Polar EF Eri during its low state of accretion, even 
though the accretion rate in MQ Dra is an order of magnitude lower than
even the low state of EF Eri. 
The similarity in phasing of the UV and optical light curves in MQ Dra 
imply a similar
location for the source of light.
We explore the possibilities of hot spots and cyclotron emission
 with simple models fit to
the UV, optical and IR light curves of MQ Dra.
To match the {\it GALEX} light curves with a single temperature circular
hot spot requires different sizes of spots for the NUV and FUV,
while a cyclotron model that can produce the optical harmonics with
a magnetic field near 60 MG 
 requires multipoles with fields $>$ 200 MG to match the UV fluxes. 

\end{abstract}


\section{Introduction}

The cataclysmic variable MQ Dra was
discovered in the Sloan Digital Sky Survey as SDSSJ155331.12+551614.5 (Szkody
 et al. 2003; S03). Its peculiar optical spectrum showed narrow, highly polarized
cyclotron emission features that could be explained with an origin from 
a white dwarf with
 a magnetic field near 60 MG. These
features underwent large periodic changes during time-resolved 
photometric and spectroscopic
observations, providing an orbital period of 4.39 hrs. Aside from the
cyclotron features, the red portion of the spectrum revealed TiO bands, indicating
an M4-M5V secondary star whose brightness
 provides a system distance of 130-180 pc (S03, Schmidt et al. 2005; S05, 
Harrison et al. 2005). 
Modeling the continuum after subtraction of the secondary indicated 
 a cool white dwarf $<$10,000K, but the flux at the blue end of the spectrum
was larger than from a white dwarf alone, suggesting an additional hot spot (S05).
X-ray observations (Szkody et al. 2004) also hinted at an upturn
at soft energies, possibly indicating a hot component. Both the appearance of
the cyclotron features and the X-ray fluxes were consistent with a cool plasma
(kT$\sim$1 keV) and extremely low accretion rate (10$^{-14}$M$_{\odot}$ yr$^{-1}$).
The exceptionally low accretion rates in these types of systems resulted in the
identification of a new category of magnetic cataclysmic variables 
called Low Accretion Rate Polars (LARPs; Schwope et al. 2002). The fact
that MQ Dra has a relatively long orbital period and a very cool white dwarf 
led to the speculation that this could be a system which is a pre-Polar i.e. the
white dwarf has never been heated by any active accretion and the cyclotron
radiation is provided only by wind accretion from the secondary. In contrast,
a Polar has known states of active accretion via a mass transfer stream
from the secondary.

This interpretation is complicated by the fact that most Polars spend much of their 
lives in low states of accretion (Ramsay et al. 2004; Araujo-Betancar et al. 2005)
that can reach low accretion rate values near those for LARPs. 
Active accretion episodes could produce heated areas that remain hot for long
periods of time after the accretion has stopped. This is borne out by UV studies
of systems during low states (Araujo-Betancar et al. 2005; G\"ansicke et al. 2006)
where 30,000-70,000K hot spots on the white dwarfs during low states were needed
to match the observed fluxes. A {\it GALEX} study of the Polar EF Eri four years after
it entered a low state (Szkody et al. 2006; S06) 
showed a 20,000K spot was needed on a 9500K white dwarf to
come close to producing the large amplitude modulations visible in the UV light curves.
However, simple black bodies for the white dwarf and spots could not reproduce both
the amplitude of the modulations and the spectral energy distributions. 
Schwope et al. (2007; Sw07) had better success with the spectral energy distribution 
using actual white dwarf models 
but they did not show UV light curves with which to compare UV observations. Since the
 white dwarfs in known active Polars observed during low states are generally hotter
than those found in LARPS, the influence of the active accretion episodes 
 is not clear. 
To obtain further insight into the heating of the white dwarf in LARPS, we undertook
{\it GALEX} observations of MQ Dra, which has a cooler white dwarf with a higher
magnetic field compared to the white dwarf in EF Eri, and an accretion rate
that is even lower than EF Eri during its lengthy low state.

\section{Observations}

The {\it GALEX} satellite obtains images with a FUV detector (1350-1750\AA) and a NUV
detector (1750-2800\AA) (Martin et al. 2005). MQ Dra was observed during two
intervals, one in June 2005 and the other in July 2006. The first set of observations
was done when the FUV detector was not operational, while both detectors were
used in 2006. The times of observation are listed in Table 1. Calibrated images
in 240s intervals were generated from the observations, phased according to the
ephemeris below and then magnitudes inside a 9 pixel radius 
aperture were measured with
the IRAF\footnote{IRAF (Image
 Reduction and Analysis
Facility) is distributed by the National Optical Astronomy Observatories, which
are operated by AURA,
Inc., under cooperative agreement with the National Science Foundation.} routine
{\it qphot}. An annulus of 3 pixels around the aperture was used to obtain and
subtract a background. The magnitudes were converted to flux from the values given
in the {\it GALEX} online documentation\footnote{See http://galexgi.gsfc.nasa.gov/tools/index.html} (FUV m$_{0}$=18.82=1.40$\times$10$^{-15}$ergs cm$^{-2}$ s$^{-1}$ \AA$^{-1}$
and NUV m$_{0}$=20.08 =2.06$\times$10$^{-16}$ ergs cm$^{-2}$ s$^{-1}$ \AA$^{-1}$). 

Since the previous available ephemeris determined from 5 nights of observations 
(S03) was
not sufficient to phase the {\it GALEX} data to a cyclotron or geometric phasing,
we obtained further optical data during 22 nights in the interval from 2007 Feb 25 - May 4.
These data were obtained on the NMSU 1m telescope at Apache Point Observatory,
using a CCD and $BVRI$ filters with integration times of 90, 60, 60, 60s
respectively. Differential light curves were constructed using
nearby stars on the same frames and calibrated with the magnitudes for
the comparison stars that were used in S03. 

As shown by the phase resolved spectra in S03 and S05, the optical light curves
of MQ Dra are dominated by the prominent cyclotron peaks at 4600, 6200 and
9200\AA, which come into the $BVRI$ bandpasses. Since these harmonics
change amplitude due to the changing viewing angle of the cyclotron region
during the orbit, a distinct sinusoidal variation is produced (S03).
Although this is a broad feature, we could combine all the nights with good 
$R$ filter coverage (16 nights) 
to
obtain a better period using the IRAF routine {\it pdm}. The past spectroscopic
and polarimetric data of S05 showed that the blue to red crossing
of the H$\alpha$ emission line was
close to
the peak amplitude of the cyclotron harmonics. Since the H$\alpha$ emission
had the same phasing and amplitude as the TiO bandheads, it allows a definition
of  phase 0 as inferior
conjunction of the secondary (and the peak of the $R$ band light curve). Thus,
we obtained the following ephemeris to phase the {\it GALEX} data:

phase 0 = HJD 2,454,156.9138 + 0.182985$\pm$0.000005E

Since the NUV light curves from 2005 and 2006 were similar, this phasing
allowed us to combine the 2005 and 2006 NUV datasets to maximize the
S/N.

Infrared photometry of MQ Dra was also obtained on the KPNO 2.1 m telescope
with the Simultaneous Quad
Infrared Imaging Device 
(SQIID)\footnote{See http//www.noao.edu/kpno/sqiid/sqiidmanual.html}, 
which obtains $JHK$ images simultaneously.
Data were obtained on two consecutive nights, 2007 May 28 and 29. On 28 May,
the first exposure began at 03:42:01 UT and data were obtained over the
next 3.3 hrs, while on May 29 the observations started at 03:31:57 and
continued for the next 2.7 hrs. The images were reduced with standard 
procedures, then
differential photometry was performed with respect to several field stars.
The differential magnitudes were
then corrected to absolute $JHK$ photometry using data from the 2MASS Point
Source Catalog, and the data were phased with the above
ephemeris.

\section{MQ Dra Light Curves}

The {\it GALEX} NUV and FUV light curves are shown in Figure 1, along with the
phased $BVRI$ light curves from 2007. All the light curves have large
variability, with peak-to-peak amplitudes of 0.7,0.8,0.5,1.0,0.9 and 0.2
mags for FUV, NUV, $B,V,R,I$ respectively.
 The $JHK$ light curves are shown in
Figure 2. The $K$-band light curve of MQ Dra reveals classic
ellipsoidal modulations with one minimum being slightly deeper than the other.
This difference can be explained if the secondary star fills (or almost
fills) its Roche lobe and is distorted so that the white dwarf-facing hemisphere
(viewed at phase $\phi$ = 0.5) is further from the center of the secondary star and therefore 
cooler.  The $H$-band
light also appears to be dominated by ellipsoidal variations, but a
non-sinusoidal light source, presumably cyclotron emission, appears to
contaminate the light curve. The $J$-band light curve is even more complex, and
only seems to bear a small imprint of the ellipsoidal variations. The spectral
type of the secondary star for MQ Dra was previously determined by S05
to be M5V. The $K$-band spectrum for this source (see Harrison et
al. 2005), however, suggests a spectral type closer to M4V due to the depth
of the Ca I triplet (at 2.26 $\mu$m). In the light curve modeling below, we
assume that the secondary star is an M4V.

The IR light curve modeling was accomplished with the Wilson-Divinney code
WD2005\footnote{WD2005 is an updated version of WD98, and can be obtained at
this website maintained by J. Kallrath:
http://josef-kallrath.orlando.co.nz/HOMEPAGE/wd2002.htm}, which includes
reflection effects. For the modeling,
we assumed a white dwarf primary with T$_{\rm eff}$ = 8,000 K, an M4V
secondary (T$_{\rm eff}$ = 3,200 K), and used the square root limb darkening
coefficients from Claret (2000b), the gravity darkening value of 0.3 (Claret
2000a) and an albedo of 0.60 (Nordlund \& Vaz 1990). The best fitting model for the $K$-band
has an orbital inclination of $i$ = 35$^{\circ}$, and is plotted in red
in Figure 2. It is clear from this figure that the light from just the
secondary star is inadequate to completely explain either the flux level or
variations seen in the $J$ and $H$-bands. Since cyclotron emission
from the $n$ = 1 (at 1.8 $\mu$m) and $n$ = 2 (at 0.9 $\mu$m) harmonics from a
B = 59 MG field is expected to be present in both the $J$ and $H$-bands, this
emission likely creates the observed discrepancy between the model and observed
light curves. 
The value for the inclination derived assuming a contact binary is much
lower than that preferred by the cyclotron modeling (i = 60$^{\circ}$, see section
6). If the assumption of a contact binary is relaxed, the infrared light
curves can be modeled with larger orbital inclinations. To investigate this
possiblity, we ran WD2005 in non-contact binary mode, but starting with a
model where the secondary star essentially filled its Roche lobe
(``R$_{contact}$''), and using stellar parameters that allowed us to reproduce
the light curves found for the contact case with i=35$^{\circ}$. We then proceeded
to shrink the radius of the secondary to create light curves for a system that
is further and further from contact. We were able to create models where
the light curves are well modeled using higher orbital inclinations. The
results are listed in Table 2. Note, however, that in this mode, the
differences in the observed minima in the model light curves are so small
as to be undetectable given the error bars on our photometry. This is true
even for the 94\% radius case.  But the observed minima $do$ differ. It is
unclear whether this is a deficiency in the light curve modeling program, or
not, but further investigation will require stronger constraints on the masses
of the two components as well as a better estimate of the temperature of the
white dwarf.

 The optical light curves are similar
in amplitude and appearance to those shown in S03 (which used arbitrary
phasing), while the increased
coverage in 2007 allows for a better resolution of the features. In particular,
the double-humped structure in the $B$ light curve is confirmed. S03 had
speculated this could be the result of cyclotron beaming with a grazing
eclipse of the accretion area at the lowest dip (phase 0.55 in Figure 1).
It is intriguing that there is a slight dip in the FUV light curve near
phase 0.05 (which is near the secondary minimum in $B$) but the poor
time resolution of the FUV and the large error bar on this single point does
not permit any conclusive link.

As in EF Eri (S06), the UV light curves of MQ Dra show a prominent 
sinusoidal modulation, with
the peak UV flux close to the peak cyclotron times, thus strongly suggesting a
link between the UV and cyclotron sources (the magnetic pole area).
 The amplitude of the modulation in the FUV in MQ Dra is
comparable to that in EF Eri, while the NUV amplitude in MQ Dra is even larger than
its FUV and about three times larger than the NUV of EF Eri. While EF Eri and
MQ Dra are known to have very different field strengths (13 vs 60 MG)
and white dwarf temperatures (9500K vs 8000K), it is obvious that a single 
temperature white dwarf (i.e. as evidenced by a constant UV light curve) 
cannot explain the observed light curves in either system.

Two possible explanations come to mind: 1) there is a hot spot on the
white dwarf near the magnetic pole causing the
UV modulation, as used by Araujo-Betancor et al. (2005) to model the
UV fluxes of low state Polars and by G\"ansicke et al. (2006) to model
the UV light curves of AM Her or 2) the UV modulation is caused by 
phase dependent cyclotron harmonics as is known to be the cause of the
optical variability. We explore these two options with simple models below.

\section{MQ Dra Spot Model}

To determine the feasibility of a hot spot, we used the
BINSYN modeling code (Linnell \& Hubeny 1996) to calculate synthetic light curves and 
spectra of MQ Dra. The model code calculates the SED of a white dwarf using
TLUSTY (Hubeny 1988) and SYNSPEC (Hubeny, Lanz \& Jeffery 1994).
The TLUSTY model uses a specified mass and radius (producing an associated
log g) and SYNSPEC produces continuum spectra plus
absorption lines of H and He.
To minimize the free parameters, we used the simplest approximations of a 
circular, isothermal hot spot to produce the light variation. 
Recognizing that satisfactory models for radiation
transfer in the presence of a strong magnetic field are not available
(Wickramasinghe \& Ferrario 2000), we used white dwarf SEDs to represent the 
spot. The BINSYN modeling code was then used to assign spot T$_{eff}$, latitude, longitude and angular
radius and, for a given parameter set, to calculate a series of synthetic 
system spectra that included contributions from the white dwarf, the spot, and
the
secondary star.
The code then integrated, for each synthetic system spectrum, the products
of the FUV, NUV and optical passbands and the synthetic system spectrum. The
resulting calculated synthetic light curves were then compared with the observed light curves.
In our model, a key element in producing the light variation with orbital phase is
the variable foreshortening and limb darkening of the spot as the white dwarf
rotates. It is important to note, in the comparison with the observations, that
neither the FUV nor NUV observed light curves are perfect sinusoids.
 
Previous estimates of the white dwarf temperature (obtained by subtracting the
secondary and cyclotron harmonics and fitting white dwarf fluxes; S03,
S05, Ferrario et al. poster at IAU Coll. 190, unpublished) obtained
temperatures from 6000-8000K with an upper limit of 10,000K. Our model uses a
temperature of 8000K. Since the Ferrario et al. model and our cyclotron  
modeling preferred inclinations near 60$^{\circ}$, and our $JHK$ light curve
modeling set a minimum inclination of 35$^{\circ}$, 
we constructed models using these two limits.

Our search for a simple spot model to simulate both the FUV and NUV
light variation of MQ Dra was unsuccessful in that
the observed light variation cannot be simulated by our model
employing a single, isothermal, circular spot which does not extend
beyond the pole. The challenge is the
approximately equal light amplitudes in FUV and NUV. Many combinations
of spot parameters can simulate either light curve individually. Table 3 lists
one pair of models for each inclination with the corresponding light
curves shown in Figure 3. For i= 60$^{\circ}$, the maximum spot angular radius
is 15$^{\circ}$ to avoid extending beyond the pole; in this case the spot center
is at a latitude of 75$^{\circ}$. The lower inclination (35$^{\circ}$) produces similar
results.

We simulated the NUV light curve 
first, and the final parameters required a spot radius close to the limiting
value. The adopted T$_{eff}$ would not produce the observed light amplitude
for a smaller spot. An
appreciably larger T$_{eff}$ would have exacerbated the difference in FUV and NUV
synthetic light amplitudes for a given common spot size. The FUV spot angular
radius was determined after the spot T$_{eff}$ had been determined, and was adjusted
to produce a synthetic FUV light amplitude matching the observed amplitude.
Note how much smaller the FUV spot radius is, as compared with the NUV spot 
radius. While a smaller FUV spot is reasonable if there is a temperature
stratification, i.e. a small hot area surrounded by a larger cool area, we
felt we did not have enough constraints within our limited data to further 
increase
the number of free parameters. 

We also extended the model fits to the optical region. As expected, the white
dwarf and spot model produce synthetic light amplitudes much smaller than 
the amplitudes observed in B or V.
This is not surprising as the optical spectra (S03,S05) show that
the optical is dominated by the cyclotron harmonics, not a stellar
continuum.

These spot models are purely empirical.	There are several options that could lead
to other solutions i.e. a non-isothermal spot with a smaller
and hotter central region contributing the bulk of the FUV source as noted
above, a non-circular
or vertically extended spot, 
 a source other than a spot (Section 6). 
Constraining these options will require better
phase-resolved UV spectra to determine the source of the light. 

\section{EF Eri Spot Model}

We also tried to model EF Eri with our latest spot model. Our original
attempt to fit the GALEX data on EF Eri (S06)
used the Wilson-Devinney code WD2005 which involves black
body representations for both the white dwarf (9800K) and the spot (20,000K).
Applying the better modeling technique described above for MQ Dra  
with
 the parameters of
EF Eri (Table 4) resulted in a spot of 24,000K with a radius of
5.5$^{\circ}$ situated at a latitude of 60$^{\circ}$.  In this case, a single size spot can
provide a reasonable match to the amplitudes of both the FUV and NUV light curves.
The fits to the UV light curves are shown in Figure 4 while the
contribution of the spot and white dwarf at the brightest phases of the lightcurve
are shown in Figure 5. As found by Sw07, using actual white dwarf
fluxes improves the match to the spectral energy distributions at maximum and minimum 
light,
but Sw07 require a larger spot than our BINSYN model fits which use
a hotter, smaller spot to produce the same amount of FUV flux. Table 5
compares the parameters of the two models. 
An important point is that while both models show that a spot model can
represent the light variation, there are enough free parameters that 
multiple solutions are possible.
A key difference between the Sw07 model and ours is that their spot encompasses
the pole and depends on the disappearance of some of the spot below the horizon
(self-eclipse) during part of an orbital cycle in order 
to produce the light variation.
Although their model produces light variation, we note that the spot is large
(half opening angle of 24$^{\circ}$) and the required spot size depends on the adopted
orbital inclination (Sw07 adopted 60$^{\circ}$, while we used 45$^{\circ}$ 
based on the Harrison et al. (2003) analysis, and Campbell et al. (2008) found
55$^{\circ}$ from their cyclotron modeling). The Sw07 model would require
an appreciably larger spot if i=45$^{\circ}$.
As in MQ Dra, it is obvious that the shape of the UV light curves is not a
perfect sinusoid, implying the hot spot is not a simple circular, isothermal spot.

\section{MQ Dra Cyclotron Model}

The cyclotron modeling was done using a Constant Lambda (CL) code first built
by Schwope et al. (1990), which uses four parameters: B (the magnetic field
strength), kT (the plasma temperature), $\Theta$ (the viewing angle to the
magnetic field), and
log$\Lambda$ (the ``size-parameter'' of the system, which is a proxy for
column density). The optical data used are the time-resolved spectra
from the Bok 2.3m that are shown in Figure 9 of S05.

To fit the models, we tried to keep the parameters as consistent with the
Ferrario et al. cyclotron model
results as possible, using B = 60 MG, kT = 1 keV, and
$i$ = 60$^{\circ}$ for initial constraints. Additionally, we used the fact
that there is a small amount of motion ($\sim$120\AA) in the $n$ = 3 (6000 \AA) harmonic to
constrain the viewing angle. Interestingly, this cyclotron
 hump appears reddest (implying
the highest value of $\Theta$) at the cyclotron minimum (phase 0.43). Coupled with the fact
that the ratio of the $n$ = 3 and $n$ = 4 harmonics show the greatest parity
at that phase, we conclude that the highest value for the viewing angle must occur at
cyclotron minimum. We initially assumed that the orbitally averaged value
of $\Theta$ would equal the system inclination, and that $\beta$ (the angle
between the rotation and magnetic axis) was the
absolute angular deviation of the viewing angle, while log$\Lambda$ was left as a free
parameter. To match the observed spectra, the models were normalized to the
data at 0.62 $\mu$m. The phase-resolved cyclotron models are shown in Figure 6,
with Table 6 listing the values of the model parameters for each spectrum.
Roughly, the orbitally averaged values for the salient parameters are: B = 59
MG, kT= 1.8 keV, and log$\Lambda$ = 3.8. The derived geometry is consistent
with a single spot $i$ = 68$^{\circ}$, and $\beta$ = 7$^{\circ}$.

While this cyclotron model does an excellent job at predicting the changing
morphology of the visual spectrum, the observed fluxes are difficult to
reconcile because for the derived geometry, the spot remains continuously
visible. With these parameters, our models suggest that the maximum amount
of flux from the cyclotron emission region should occur at $\phi$ = 0.43, but
this is near where the observed $minimum$ occurs. Given that the n = 3 and n = 4
harmonics have nearly identical fluxes, and are reddest at this phase,
the viewing angle should be at a maximum, which in turn should produce the
greatest amount of observed cyclotron emission. Comparing the raw,
un-normalized model intensities to the photometry suggests that the fluxes
at photometric minimum are down by a factor of six if we assume an equal
emitting area at all times.  We can explain this variation if we assume the
values for $i$ and $\beta$ derived from the cyclotron modeling are incorrect,
and that the accretion region is partially self-eclipsed. 

For a simple geometry (a uniform, cylindrical column), we find that we can
relate $i$ + $\beta$ to the column height (h) through the following equation:

h = (1 + $\eta$)R$_{\rm WD}$(tan$^{2}$($i$ + $\beta$ - 90) - 1.0)

where $\eta$ is the fraction of the column that remains visible, and
R$_{\rm WD}$ is the radius of the white dwarf. Assuming values of 0.01 $\leq$
h $\leq$ 0.05 R$_{\rm WD}$, we found values of 97.5$^{\circ}$ $\leq$ ($i$ +
$\beta$) $\leq$ 106.5$^{\circ}$, as listed in Table 7. The geometries implied by
these combinations of $i$ and $\beta$ cannot reproduce the observed viewing
angles without curvature of the field lines. Assuming a dipole field,
Beuermann, Stella \& Patterson (1987) show that the angle of curvature, $b$, is
related to the magnetic colatitude by the equation $b$ = 3/2 $\beta$. We used
a root-finding scheme to find values of $b$ for a range of orbital
inclinations that reproduce the observed phase dependence of the viewing angle
$\theta$ (see Fig. 7). The mean value for $b$ is 68$^{\circ}$, which implies
$\langle$ $\beta$ $\rangle$ $\sim$ 45$^{\circ}$. This, in turn, restricts
the orbital inclination to the values listed in Table 7. We believe that
this is the simplest interpretation, but more complex field structures
and/or accretion regions could also reproduce the observations.

Finally, we attempted to fit the observed SED at both photometric minimum and maximum.
In Figure 8, we show the observed photometry for these two phases,
obtained from the GALEX, NMSU, and KPNO data for the UV, optical, and
near IR, respectively. Using the color-magnitude table for pure hydrogen
white dwarfs with log($g$) = 8 in Bergeron, Wesemael, and Beauchamp (1995)
at 180 pc with $\lambda F_{\lambda}\simeq$ 20\% below the NUV photometry
at photometric minimum implies T $\simeq$ 8400 K. In this work, we adopt
a slightly lower temperature of 8000 K. This WD (green line) is  co-added
with a M4 secondary normalized to the K-band  photometry. In addition, we
added our optical cyclotron models for both phases, finding that the SEDs were
well reproduced in the IR and the optical, but that the noted UV photometric
variability was still unexplained. Therefore, we added a second cyclotron model to
explain the UV photometry. As in EF Eri (Campbell et al. 2008), 
it was difficult to constrain the UV cyclotron models without
proper phase-resolved UV spectroscopy. We assumed 
that the UV cyclotron model has identical model parameters to the
optical model, but with the magnetic field strength increased to push the
harmonics into the UV. A magnetic field of B =  235 MG produced harmonics in the
right places; this value was the lowest field strength
which could be reasonably fit to the observed data. This high field strength is not
unknown among Polars as AR UMa shows a similar value for B (Schmidt et al. 1996). To calibrate the fluxes, we
assumed that the peak flux (in $F_{\lambda}$) of the $n$ = 3 harmonics for both
models were identical. The model spectrum was then integrated over the
transmission function of the GALEX NUV and FUV bandpasses using the IRAF
package CALCPHOT, and the synthetic photometric point compared to the GALEX
photometry. 

At photometric maximum, we found that while the NUV point was well
predicted, the model could not produce enough flux in the FUV. Thus, we
increased log$\Lambda$ in that model from 3.6 to 4.0, which then adequately
explained the fluxes in that bandpass. At photometric minimum, however, we
found that the predicted flux was well down (by a factor of 5) from the
GALEX data. While the nature of this discrepancy remains unclear, it is
possible that the UV cyclotron emission region is not co-located
with the optical accretion spot. If it were located closer to
the pole, the accretion spot would just barely self eclipse, leaving a much
larger emitting surface area and that could possibly explain this result.

\section{Discussion}

As UV data accumulate on Polars during low states of accretion, it is
increasingly obvious that some areas emitting substantial UV flux remain on the
white dwarf even
during states of very low accretion when there is no evidence of a mass
transfer stream. Studies of Polars at random times (Ramsay et al. 2004;
Araujo-Betancor et al. 2005) show that many Polars are in low states, implying
they spend much of their lifetime in these states, but the length of these
low states are not usually known and the data that do exist reveal the low 
states vary in length. For a few well-studied systems, the extra
UV light over that of the white dwarf continuum 
is known as a function of time since the onset of the low state and is
modelled with a hot spot. The
study of AM Her (G\"ansicke et al. 2006) with HST and FUSE for several months
starting 200 days after it began a low state showed that the spot was
about 10,000K cooler than its high state fitted value of 47,000K, but of a 
similar size as compared to a high state of accretion.
There was also a slight change in latitude of the spot between the two states.
These characteristics did not change during the 4 months of the low state.
However, AM Her is known to have frequent high states of accretion which can
heat up a spot and the derived accretion rate during low states is fairly high at
6$\times$10$^{-12}$M$_{\odot}$ yr$^{-1}$. This value is about a factor of 10
below the typical mass accretion rate for a Polar in a high accretion state
but a factor of 100 larger than a typical LARP (S05). The white dwarf temperature at 20,000K
is close to the value of typical cataclysmic variables that are heated by
long term accretion (Sion 1999, Townsley \& Bildsten 2004). 

Our UV (S06) and the
X-ray (Sw07) study of EF Eri probe a longer and lower state of accretion
compared to AM Her. The
GALEX and X-ray data were obtained 7 and 5 yrs respectively into the low state 
that began in 1997 and has now reached 10 yrs in length. SW07
estimate the accretion luminosity during the low state as 2.4$\times$10$^{30}$
erg s$^{-1}$, which corresponds to an accretion rate of about 
3$\times$10$^{-13}$M$_{\odot}$ yr$^{-1}$. The corresponding specific accretion 
rate of 0.01 g cm$^{-2}$s$^{-1}$
puts EF Eri into the bombardment regime where there is no shock at the accretion
column and there is only cyclotron cooling in the atmosphere of the white dwarf
which result in the observed harmonics in the near-IR (Kuijpers \& Pringle 1982;
Woelk \& Beuermann 1996; Wickramasinghe \& Ferrario 2000). The resulting lower
temperature of the white dwarf in EF Eri than in AM Her is consistent with low
accretion values and long time intervals spent at low states.

MQ Dra is the most extreme case studied so far. It has one of the coolest
white dwarfs and one of the lowest accretion rates 
 (6$\times$10$^{-14}$M$_{\odot}$ yr$^{-1}$; S05) known among close binaries
containing magnetic white dwarfs.
It has never been seen in a high accretion state and the low
white dwarf temperature and long orbital period indicate it may never have been
in an active state. Thus, it is surprising that even in this regime, there is
a significant UV source of light above that from a white dwarf that fits
the optical continuum between cyclotron harmonics, and this source is highly
variable during the orbit.
If this UV source is from a heated area around the magnetic pole, the capture of the stellar wind from the secondary by the magnetic
white dwarf must be able to produce a significant heating of the atmosphere for
extended times. As noted by S05, Li et al. (1994;1995) determined that for
fields near 50-100MG, the magnetic white dwarf can capture all of the
stellar wind from the secondary, thus increasing the amount coming to
the white dwarf. The figures in these papers suggest that the wind still
is confined to a footprint near the magnetic pole(s). While the exact shape
of a heated spot is not clear in this situation, it is reasonable that
the heated area would be in the same location as the pole and in phase
with the optical cyclotron emission (the low optical depth ensures the 
lowest harmonics are optically thin and hence the cyclotron is not
beamed as in the optically thick case). In addition, the magnetic coupling between the two stars
can lead to increased stellar activity (flares and coronal mass ejections)
on the secondary, thus increasing the accretion rate sporadically and
creating more heating (Howell et al. 2006). However, a simple hot spot is not an ideal fit to
the observations as
described above. 

An alternate possibility is that the extra UV flux and modulation is
caused by cyclotron harmonics in the UV. 
The greatest
problem with this interpretation is that to have prominent harmonics in
the UV requires a magnetic field $>$ 200 MG. Possible harmonics are evident
in the UV in AR UMa (G\"ansicke et al. 2001) which is known to have
a field of about 240 MG. However, recent modeling of the optical flux
and polarization of EF Eri (Beuermann et al. 2007) shows that this system
(as well as other Polars) has complex
fields, with multipoles that could include fields up to 110MG. Recent success
in modeling IR phase-resolved spectra of EF Eri (Campbell et al. 2008)
was achieved with a two component cyclotron model. The extension of this type of model
to the UV with a field of 115 MG can explain the amplitudes of
the {\it GALEX} light curves for EF Eri. If MQ Dra has such a complex multipole field,
then the optical cyclotron could be produced by fields near 60 MG and
the UV cyclotron by the components near 200 MG. The correct interpretation
could be resolved by time-resolved UV spectra which would reveal whether
these cyclotron harmonics actually exist in the UV.

\section{Conclusions}

Our GALEX NUV and FUV light curves have revealed that the lowest accretion
rate Polars, with no active stream accretion, still show large amplitude
periodic UV variability with similar phasing and amplitude to the optical
light curves. The extreme
case of MQ Dra, with a long orbital period and a white dwarf of only 8000K,
shows similar FUV variability to that of EF Eri, a Polar with extended low states
but also a known high state of accretion in the past. 
The sequence of white dwarf temperatures appears to
follow the accretion rates, with the lowest temperature white dwarfs existing
in the systems with the lowest
accretion, implying some long term heating effects even in these cases of
wind-accretion. 
However, attempts to fit the UV light curves with a simple isothermal hot spot 
cannot adequately
describe the observations. While this simple type of spot can produce the amplitudes
of the NUV and FUV modulation in EF Eri,
it cannot reproduce the shape of
the light curves well, implying distorted, multi-temperature spots or additional
components. MQ Dra is harder to fit with a spot than EF Eri, as the similar amplitudes in NUV and
FUV cannot be reproduced with a similar size spot at a given temperature.

Another explanation of the UV phasing and amplitudes might be possible
using cyclotron harmonics, but only if the white dwarf has multipoles with
fields extending above 200 MG.
While we could adequately model optical phase-resolved spectroscopy of MQ Dra
with a cyclotron model having B = 59 MG, kT $\simeq$ 2.0 keV, and log$\Lambda$ =
4.0, both the observed motion of the
harmonics and the morphological appearance of the spectra at cyclotron minimum
and maximum defied our simplistic expectations. To explain these results
in a consistent way would require a secondary that does not fill its
Roche lobe, an accretion column that has a high magnetic co-latitude 
($\beta$ = 45$^{\circ}$), field lines that are tilted relative to the local normal,
and an emitting region of the white
dwarf that decreases by a factor of 6 from cyclotron maximum to minimum.
The SED from the FUV to the $K$-band can be fit with
an 8000 K WD, an M4 secondary normalized to the $K$-magnitude at
photometric minimum and two cyclotron models (our derived optical
cyclotron model and a second analogous cyclotron model with an equivalent peak
flux in $n$ = 3). This combination can reproduce the observed UV photometric 
variability, but
at photometric minimum the UV cyclotron model must be scaled by an arbitrary
value of 5 (possibly explained by moving the UV accretion spot
closer to the rotation axis).

 UV phase resolved spectra are needed to determine which model to pursue
in more detail. The spectra would enable the temperature structure of a
hot spot to be determined and reveal if cyclotron features from high
fields are present. 
Further theoretical models of the atmospheric heating and cooling in the
bombardment case of very low specific accretion would also help to
determine the nature of the heated areas surrounding the magnetic poles
at these low rates.
 
\acknowledgments

Support for this research was provided by NASA GALEX grant NNG05GG46G.
We thank Hugh Harris for providing standard stars for the calibration
of the optical data and Gary Schmidt for the use of his optical spectra
of MQ Dra for the cyclotron model and for his comments on our models.

\clearpage
\begin{deluxetable}{ccl}
\tablewidth{0pt}
\tablecaption{Summary of Observations}
\tablehead{
\colhead{Date} & \colhead{Time} & \colhead{Data}}
\startdata
20050607 & 04:48-19:52 & GALEX NUV 10 visits\\
20050723 & 11:53-12:13 & GALEX NUV 1 visit \\
20060715 & 16:40-17:04 & GALEX NUV,FUV 1 visit \\
20060716 & 01:08-02:55 & GALEX NUV,FUV 2 visits \\
20060716 & 07:27-09:30 & GALEX NUV,FUV 2 visits \\
20060716 & 17:20-21:02 & GALEX NUV,FUV 3 visits \\
20060717 & 01:46-03:35 & GALEX NUV,FUV 2 visits \\
20060722 & 03:13-03:33 & GALEX NUV,FUV 1 visit \\
\enddata
\end{deluxetable}

\clearpage
\begin{deluxetable}{cc}
\tablewidth{0pt}
\tablecaption{Roche-filling Factor vs Inclination}
\tablehead{
\colhead{R/R$_{contact}$} & \colhead{i$^{\circ}$} }
\startdata
1.00 & 35 \\
0.94 & 40 \\
0.90 & 47 \\
0.86 & 54 \\
0.82 & 65 \\
\enddata
\end{deluxetable}

\clearpage
\begin{deluxetable}{lcc}
\tablewidth{0pt}
\tablecaption{MQ Dra Model Parameters}
\tablehead{
\colhead{Parameter} & \colhead{i=60$^{\circ}$} & \colhead{i=35$^{\circ}$}}
\startdata
WD T$_{eff}$  & 8000K & 8000K \\
WD mass & 0.6M$_{\odot}$  & 0.6M$_{\odot}$ \\
WD radius & 0.012R$_{\odot}$ & 0.012R$_{\odot}$ \\
WD log g &  8.0 & 8.0 \\
Spot T$_{eff}$  & 20,000K & 24,000K \\
Spot latitude & 73$^{\circ}$ & 65$^{\circ}$ \\
Spot ang rad &  5$^{\circ}$ (FUV) & 3.5$^{\circ}$ (FUV) \\
Spot ang rad & 12$^{\circ}$ (NUV) & 14$^{\circ}$ (NUV) \\
\enddata
\end{deluxetable}

\clearpage
\begin{deluxetable}{lc}
\tablewidth{0pt}
\tablecaption{EF Eri Model Parameters}
\tablehead{
\colhead{Parameter} & \colhead{Value}}
\startdata
WD T$_{eff}$  & 9500K \\
WD mass & 0.6M$_{\odot}$ \\
WD radius & 0.012R$_{\odot}$ \\
WD log g &  8.0 \\
Orbit i &   45$^{\circ}$ \\
Spot T$_{eff}$  & 24,000K \\
Spot latitude & 60$^{\circ}$ \\
Spot ang rad &  5.5$^{\circ}$ \\
\enddata
\end{deluxetable}

\clearpage
\begin{deluxetable}{lcc}
\tablewidth{0pt}
\tablecaption{Comparison of EF Eri Models}
\tablehead{
\colhead{Parameter} & \colhead{Sw07} & \colhead{This work}}
\startdata
WD T(K) & 9750 & 9500 \\ 
Spot T (K) & 18,500 & 24,000 \\
Inclination ($^{\circ}$) & 60 & 45 \\
Spot Ang  radius ($^{\circ}$) & 24 & 5.5 \\
Spot Latitude ($^{\circ}$) & 77.5 & 60 \\
\enddata
\end{deluxetable}

\clearpage
\begin{deluxetable}{cccclc}
\tablecaption{Cyclotron Modeling Parameters for MQ Dra}
\tablewidth{0pt}
\tablehead{
\colhead{Spectrum} & \colhead{Phase} & \colhead{B (MG)} & \colhead{T (keV)} &
\colhead{$\Theta$} &\colhead {$Log(\Lambda)$} }
\startdata
1  & 0.67 & 59.0 & 2.0 & 70.0 & 4.1\\
2  & 0.73 & 59.0 & 1.6 & 68.0 & 4.1\\
3  & 0.82 & 59.0 & 1.6 & 65.0 & 3.8\\
4  & 0.89 & 59.0 & 1.8 & 63.0 & 3.6\\
5  & 0.95 & 59.0 & 1.8 & 61.0 & 3.6\\
6  & 0.02 & 59.0 & 1.8 & 63.0 & 3.6\\
7  & 0.09 & 59.0 & 1.8 & 65.0 & 3.6\\
8  & 0.16 & 59.0 & 1.8 & 67.0 & 3.7\\
9  & 0.23 & 59.0 & 1.8 & 70.0 & 3.7\\
10 & 0.30 & 59.0 & 1.9 & 72.0 & 3.7\\
11 & 0.36 & 59.0 & 2.0 & 74.0 & 3.9\\
12 & 0.43 & 59.0 & 2.0 & 76.0 & 3.9\\
13 & 0.50 & 59.0 & 2.0 & 74.0 & 3.9\\
14 & 0.57 & 59.0 & 2.0 & 72.0 & 3.9\\
\enddata
\end{deluxetable}

\clearpage
\begin{deluxetable}{ccc}
\tablewidth{0pt}
\tablecaption{Relation of Column Height to $i$ and $\beta$}
\tablehead{
\colhead{h$_{col}$ (R$_{WD}$)} & \colhead{$i$ + $\beta$} & \colhead{$i$} }

\startdata
0.01 & 97.5$^{\circ}$ & 52.5$^{\circ}$ \\
0.02 & 100.5$^{\circ}$ & 56.5$^{\circ}$ \\
0.03 & 102.9$^{\circ}$ & 57.9$^{\circ}$ \\
0.04 & 104.8$^{\circ}$ & 59.8$^{\circ}$ \\
0.05 & 106.5$^{\circ}$ & 61.5$^{\circ}$ \\
\enddata
\end{deluxetable}

\clearpage
\begin{figure} [h]
\figurenum {1}
\plotone{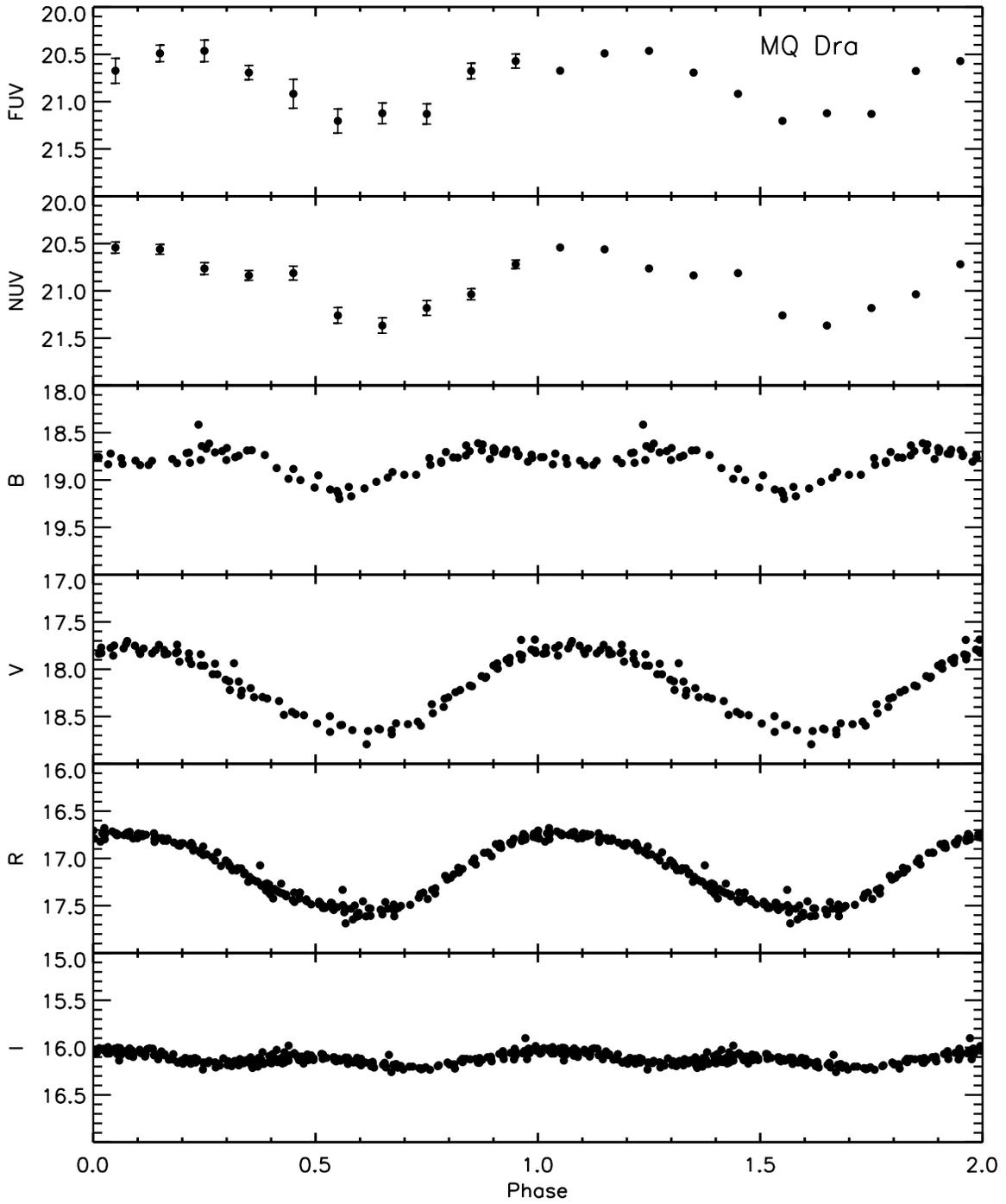}
\caption{GALEX FUV and NUV and optical B,V,R,I filter light curves as a 
function of phase (phase 0 is cyclotron max and inferior conjunction of 
secondary). Light curves are repeated from phases 1.0 to 2.0, with error bars shown only
for the first cycle. Error bars for the optical data are smaller
than the points.}
\end{figure}

\clearpage
\begin{figure}
\figurenum {2}
\epsscale{0.8}
\plotone{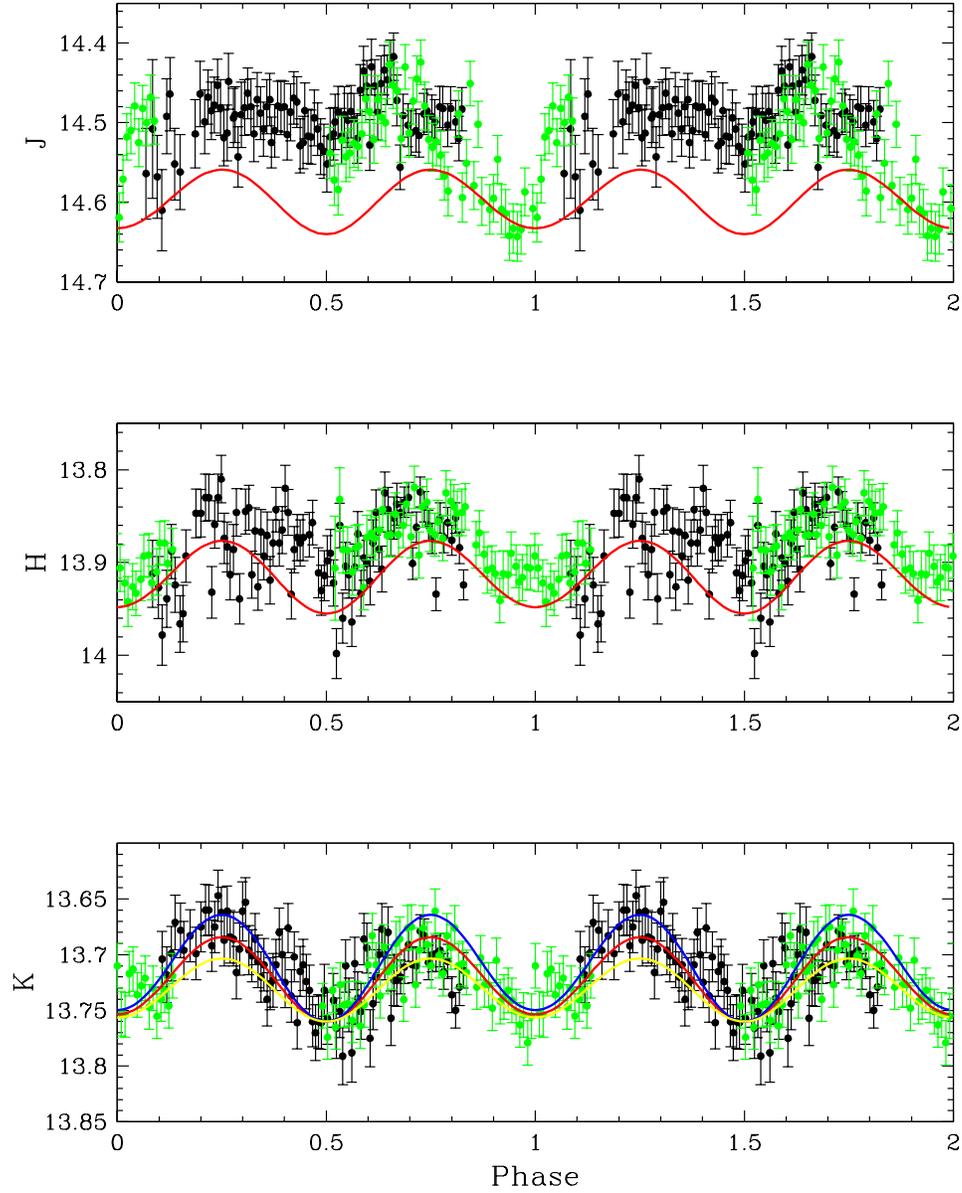}
\caption{$J,H,K$ light curves of MQ Dra (green is May 28 and black is May 29) 
fit with a model of an M4 secondary filling its Roche lobe
and an 8000K white dwarf. Inclinations of 40, 35 and 30$^{\circ}$ are shown as
blue, red and yellow lines respectively; for clarity, only the best fitting
35$^{\circ}$ fit is shown for J,H.}
\end{figure}

\clearpage
\begin{figure}
\figurenum {3}
\epsscale{0.9}
\plotone{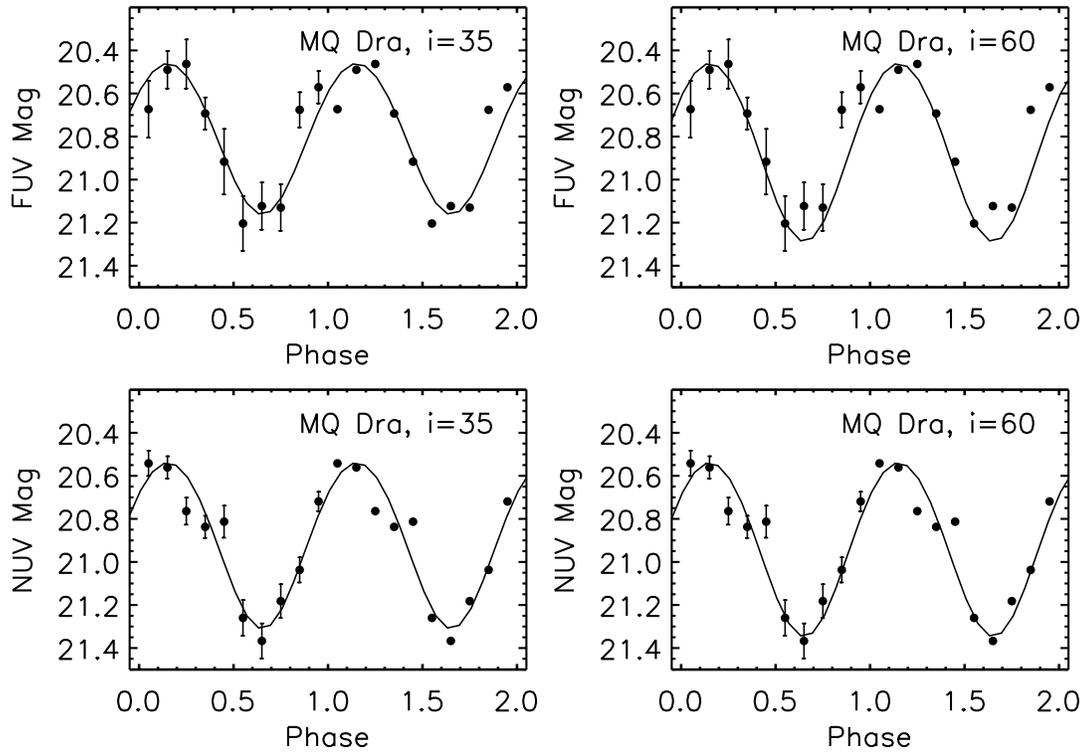}
\caption{Spot Model fits from Table 3 for the GALEX FUV (top) and NUV (bottom)
light curves of MQ Dra for
inclinations of 35$^{\circ}$ (left) and 60$^{\circ}$ (right).} 
\end{figure}

\clearpage
\begin{figure}
\figurenum {4}
\plotone{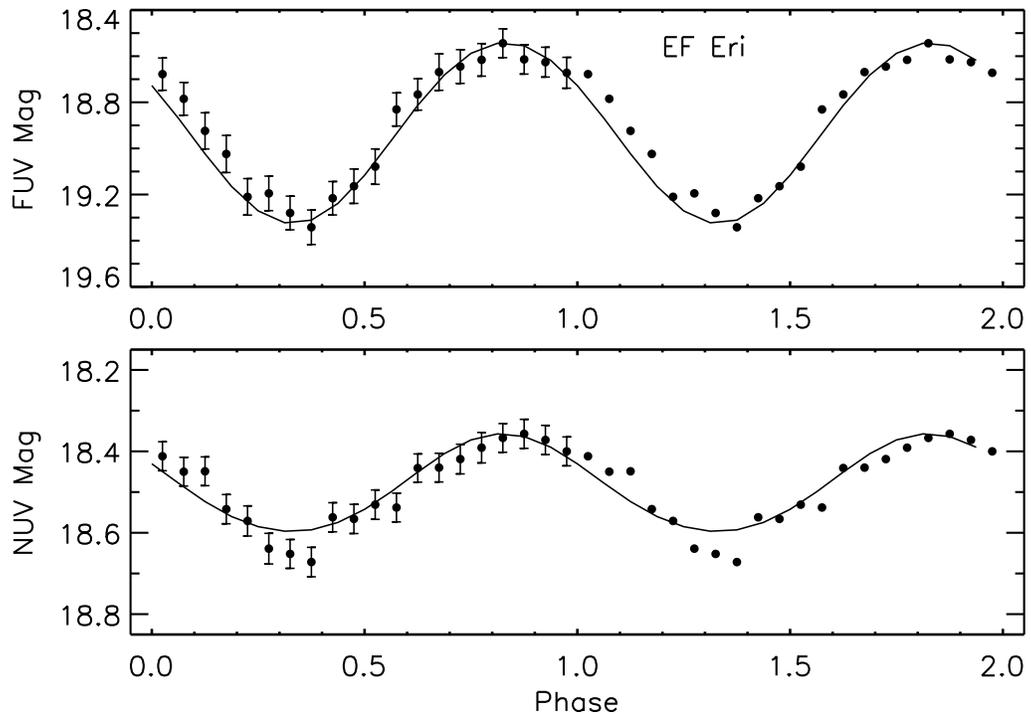}
\caption{Model fit for EF Eri with a 24,000K spot at 60$^{\circ}$ latitude and 5.5$^{\circ}$ radius
 on a 9500K white dwarf.}
\end{figure}

\clearpage
\begin{figure}
\figurenum {5}
\plotone{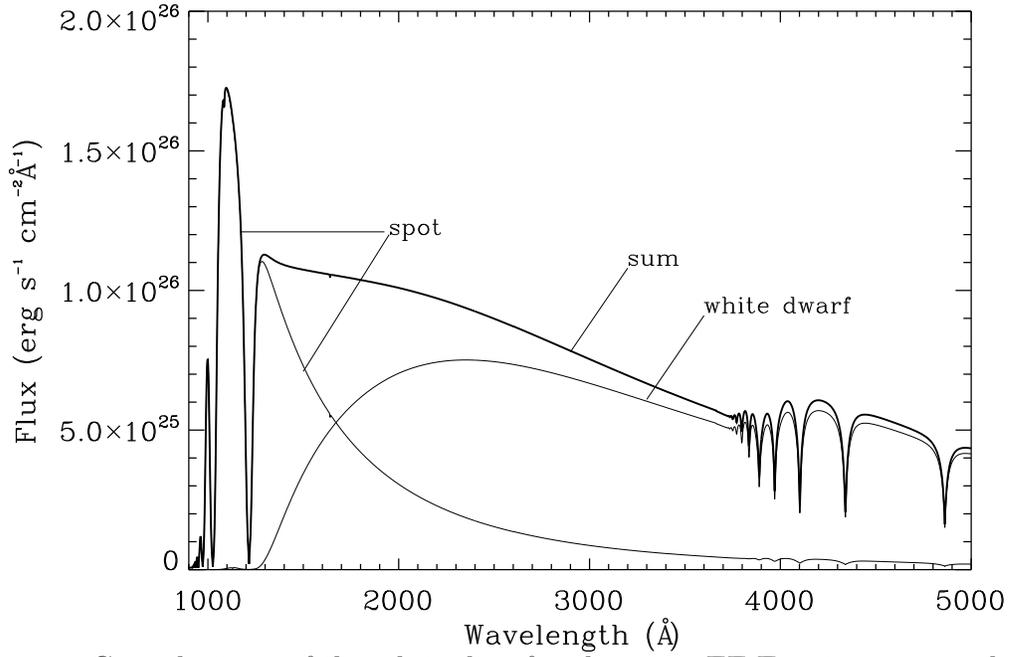}
\caption{Contributions of the white dwarf and spot in EF Eri at maximum light during the orbit.}
\end{figure}

\clearpage
\begin{figure}
\figurenum {6}
\epsscale{0.80}
\plotone{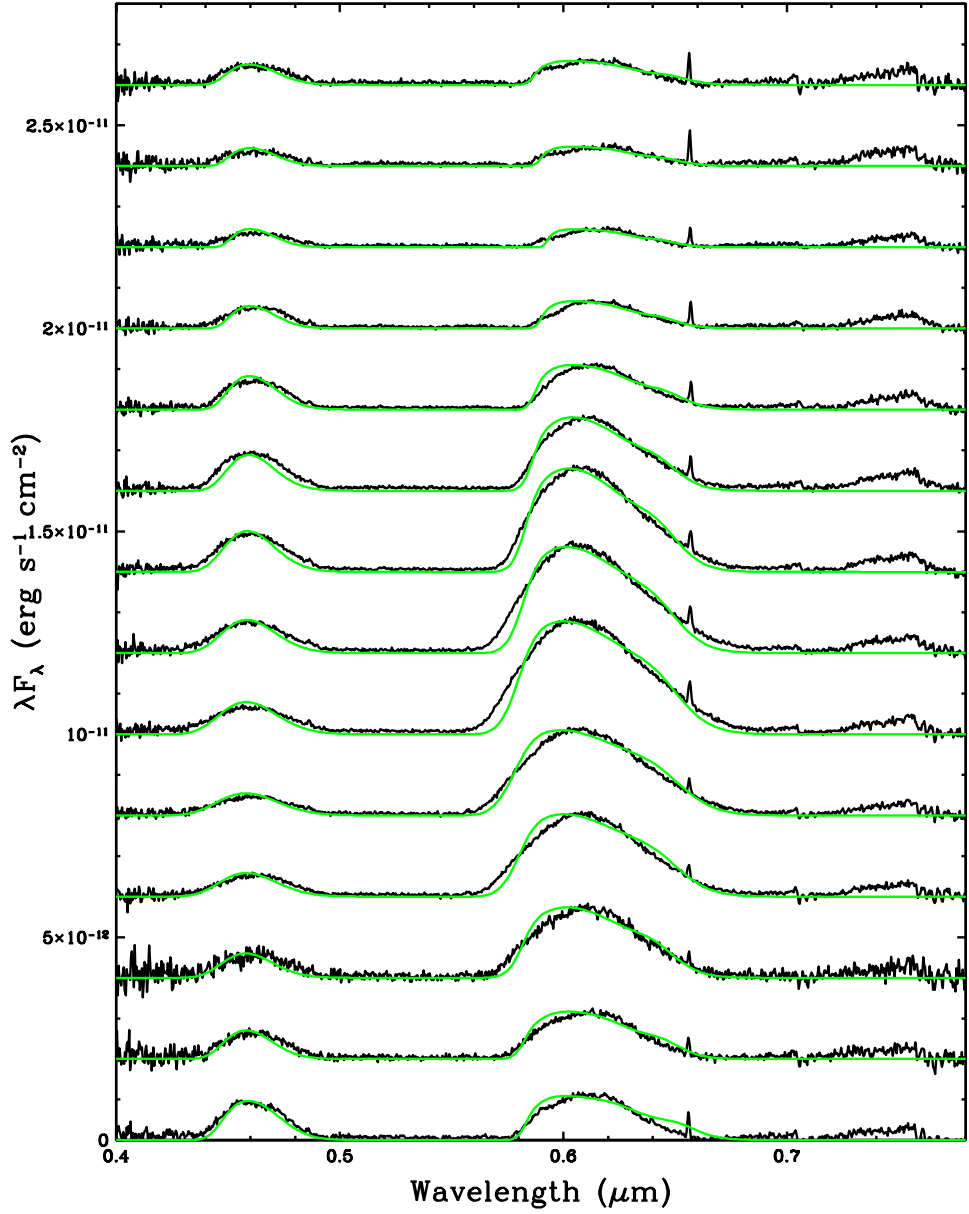}
\caption{Phase-resolved spectra of MQ Dra, data (in black) from S05 are
 stacked with a constant increment of 2.0 $\times$ 10$^{-12}$
erg s$^{-1}$ cm$^{-2}$. The best fit cyclotron models (in green) are
fit over the cyclotron models, with spectrum 1 at bottom and 14 at top from
Table 6.}
\end{figure}

\clearpage
\begin{figure}
\figurenum {7}
\epsscale{0.60}
\plotone{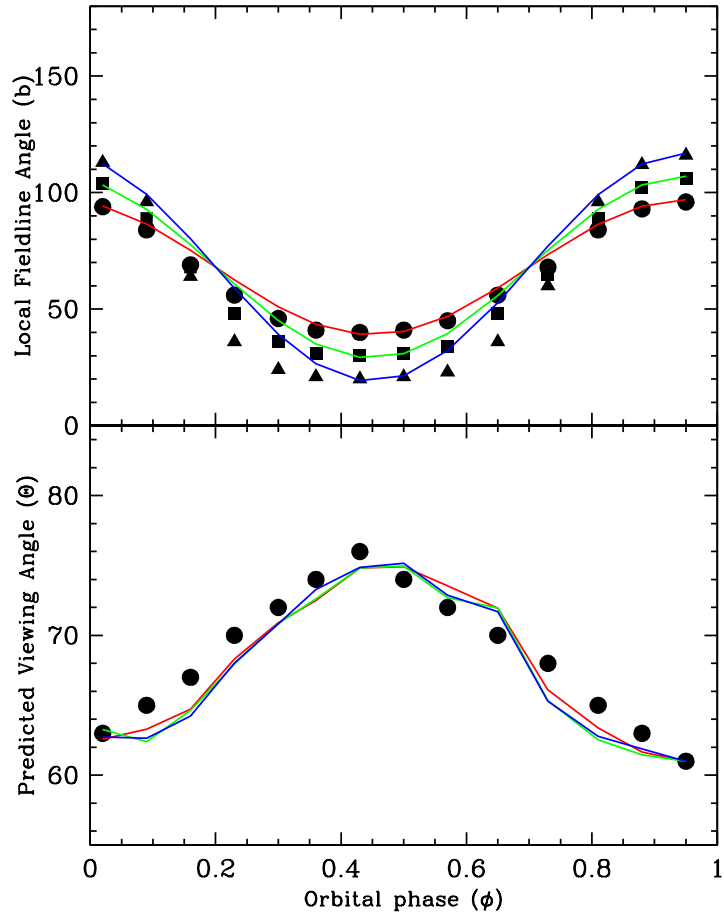}
\caption{ (top) The change in the local field angle for three orbital 
inclinations 35$^{\circ}$ (circles), 45$^{\circ}$ (squares), 55$^{\circ}$ (triangles), that
reproduced the observed viewing angle variation (bottom panel, circles).
The colored lines shown in the top panel to guide the eye are reproduced in
the bottom panel to demonstrate that this interpretation can reproduce
the observed variation in viewing angle.
}
\end{figure}

\clearpage
\begin{figure}
\figurenum {8}
\epsscale{1.0}
\plottwo{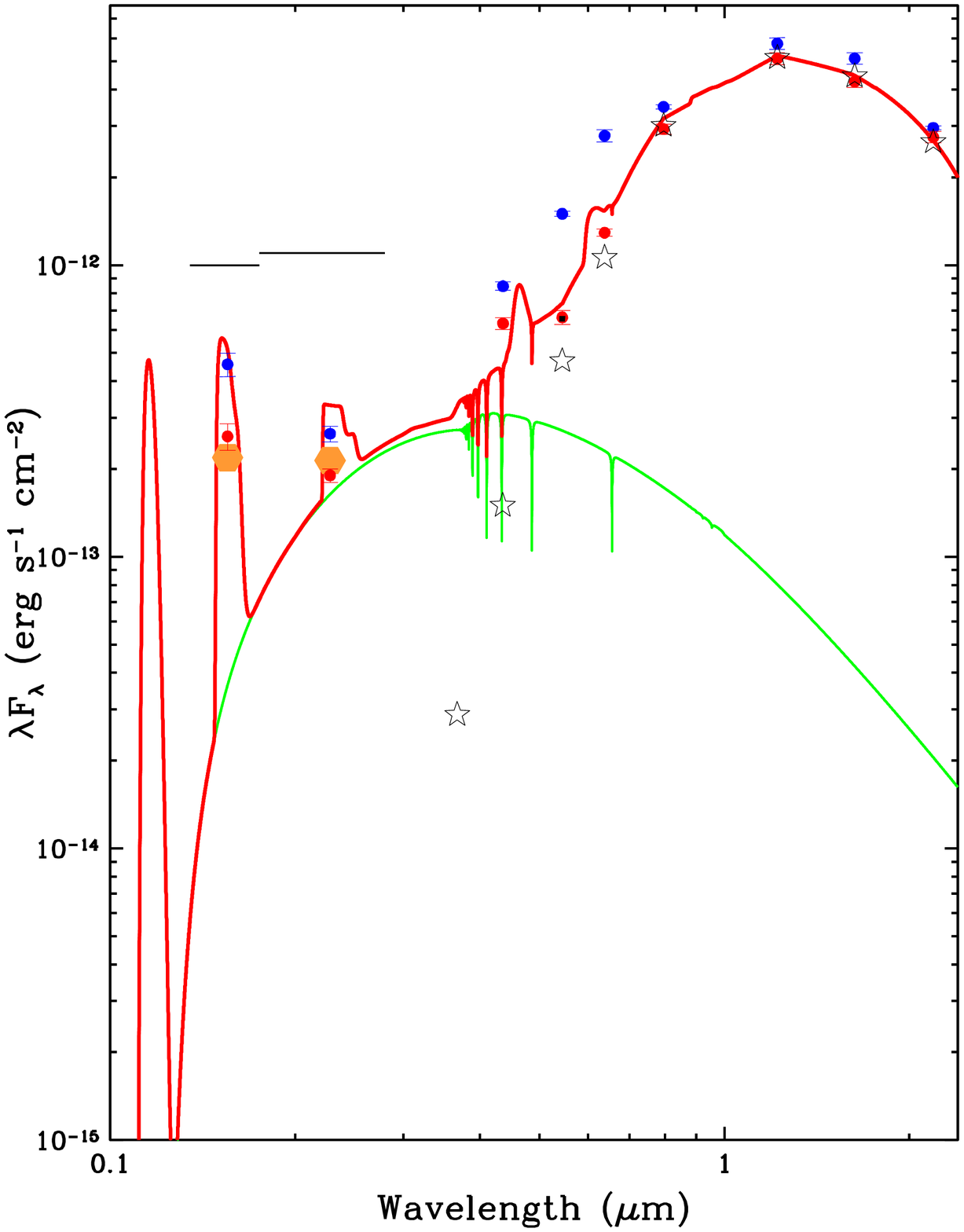}{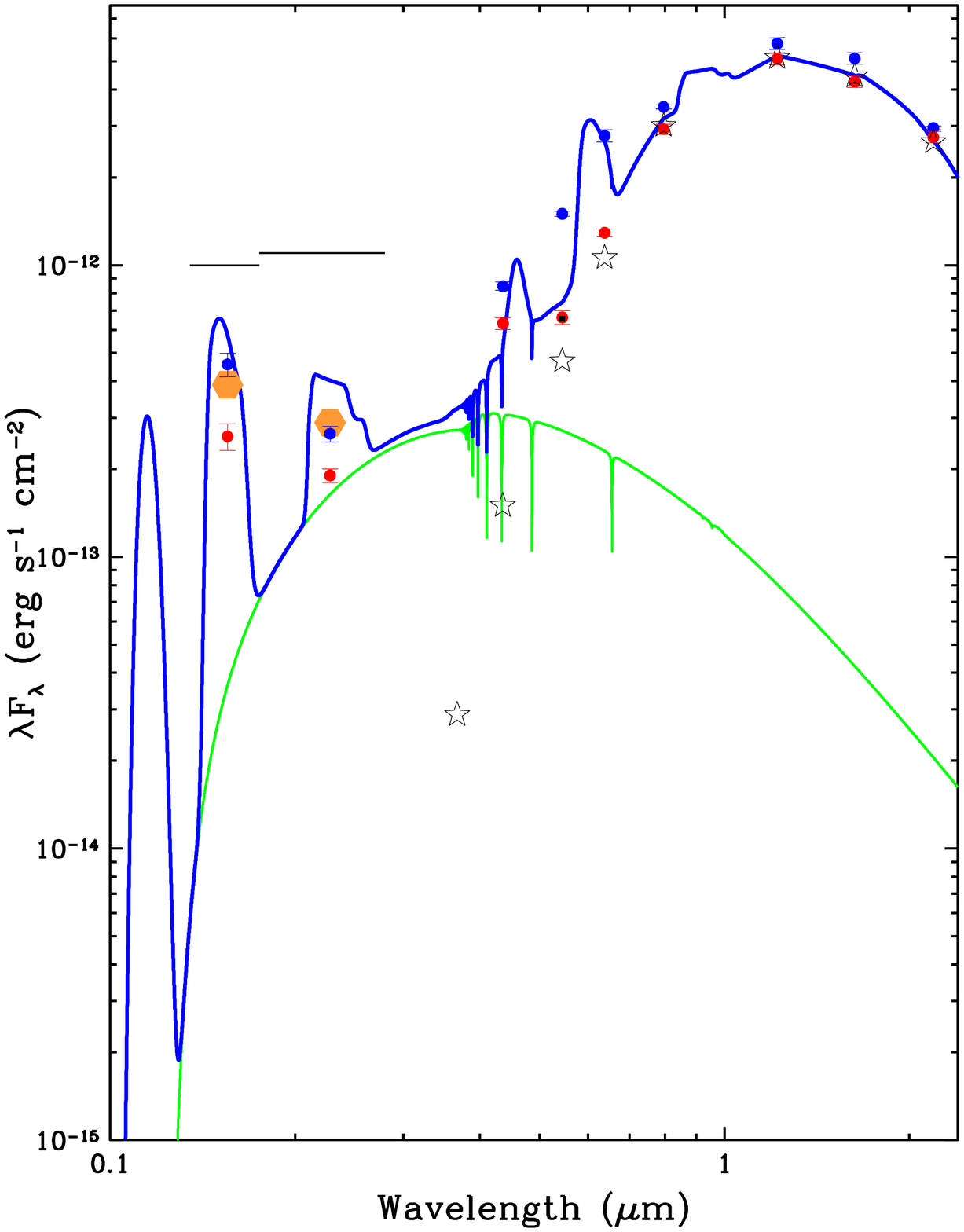}
\caption{(a) Cyclotron model SED of MQ Dra at photometric minimum ($\phi$ = 0.65). Horizontal
lines are GALEX bandpasses, points are UV
optical and IR photometric observations (red are photometric min, blue are
photometric max), and the green line is the spectrum of an 8000K
WD, normalized to $\simeq$ 20\% below the NUV photometric point. Stars are an M4
spectrum normalized to the $K$-band point. Our cyclotron model at photometric
minimum and an analogous UV model, with B = 235 MG, are added in normalized so
that the 
peak-value of $F_{\lambda}$ in the $n$ = 3
harmonics of both the optical and UV cyclotron models are identical, but 
the UV model is scaled up by a factor of five (see text for details). 
The overall fit is plotted in red. 
Orange hexagons are the NUV and FUV photometry implied by our models assuming
the NUV and FUV transmissions. (b)
Similar but for photometric max, with the final
composite model now shown in blue. No 
scaling to account for the UV fluxes is used in this case.}
\end{figure}


\begin{references}

\reference{} Araujo-Betancor, S. et al. 2005, \apj, 622, 589
\reference{} Bergeron, P., Wesemael, F. \& Beauchamp, A. \pasp, 107, 1047
\reference{} Beuermann, K., Stella,L. \& Patterson, J., 1987, \apj, 316, 360 
\reference{} Beuermann, K., Euchner, F., Reinsch, K., Jordan, S. \& G\"ansicke, B. T. 2007, \aap, 463, 647
\reference{} Campbell, R. K., Harrison, T. E., Schwope, A. \& Howell, S. B. 2008, \apj, 672, 531
\reference{} Claret, A. 2000a, \aap, 359, 289
\reference{} Claret, A. 2000b, \aap, 363, 1081
\reference{} G\"ansicke, B. T., Long, K. S., Barstow, M. A. \& Hubeny, I. 2006, \apj, 639, 1039
\reference{} G\"ansicke, B. T., Schmidt, G. D., Jordan, S. \& Szkody, P. 2001, \apj, 555, 380
\reference{} Harrison, T. E., Howell, S. B., Szkody, P. \& Cordova, F. A. 2005, \apj, 632, L123
\reference{} Howell, S. B., et al. 2006, \apj, 652, 709
\reference{} Hubeny, I. 1988, Comp. Phys. Comm., 52, 103
\reference{} Hubeny, I., Lanz, T., \& Jeffery, C.~S. 1994, in Newsletter on
Analysis of Astronomical Spectra No. 20, ed. C. S. Jeffery (CCP7;St. Andrews: St. Andrews Univ.), 30
\reference{} Kuijpers, J. \& Pringle, J. E. 1982, \aap, 114, L4
\reference{} Li, J., Wickramasinghe, D. T., \& Wu, K. W. 1995, \mnras, 276, 255
\reference{} Li, J., Wu, K. W. \& Wickramasinghe, D. T. 1994, \mnras, 268, 61
\reference{} Linnell, A. P. \& Hubeny, I. 1996, \apj, 471, 958
\reference{} Martin, D. C. et al. 2005, \apj, 619, L1
\reference{} Nordlund, A. \& Vaz, L. P. R. 1990, \aap, 228, 231
\reference{} Ramsay, G. et al. 2004, \mnras, 350, 1373
\reference{} Schmidt, G. D. et al. 1996, \apj, 473, 483
\reference{} Schmidt, G. D. et al. 2005, \apj, 630, 1037 (S05)
\reference{} Schwope, A. D. 1990, Reviews in
Modern Astronomy, 3, 44
\reference{} Schwope, A. D. et al. 2002, in ASP Conf. Vol 261, The Physics of
Cataclysmic Variables and Related Objects, ed. B. T. G\"ansicke, K. Beuermann, \& K. Reinsch (San Francisco:ASP), 102
\reference{} Schwope, A. D., Staude, A., Koester, D. \& Vogel, J. 2007, aap, 469, 1027 (Sw07)
\reference{} Sion, E. M. 1999, \pasp, 111, 532
\reference{} Szkody, P. et al. 2003, \apj, 583, 902 (S03)
\reference{} Szkody, P. et al. 2004, \aj, 128, 2443
\reference{} Szkody, P. et al. 2006, \apj, 646, L147 (S06)
\reference{} Townsley, D. M. \& Bildsten, L. 2004, \apj, 600, 390
\reference{} Wickramasinghe, D. T. \& Ferrario, L. 2000, \pasp, 112, 873
\reference{} Woelk, U. \& Beuermann, K. 1996, \aap, 306, 232
\end{references}
\end{document}